\documentclass[aps,showpacs,superscriptadress,preprint]{revtex4}
\usepackage{graphicx}

\begin{document}
\title{A possible explanation for AMS doubly charged anomalous event}
\author{Yun Zhang$^1$, and Ru-Keng Su$^2$}
\affiliation{$^1$Department of physics, Fudan University, Shanghai 200433 , P. R. China\\
$^2$China Center of advanced Science and Technology (World Laboratory)\\
P. O. Box 8730, Beijing 100080, P. R. China \\}

\begin{abstract}
By means of the quark mass density-dependent model, a possible explanation
for the doubly charged anomalous event with $\frac ZA=0.114$ reported by
Alpha Magnetic Spectrometer Collaboration is given. It seems a strangelet.
The composition, radius and mean lifetime of this strangelet are given.
\end{abstract}

\maketitle

According to the argument of Greiner et. al \cite{Greiner1}, the strangelet
could serve as an unambiguous signature for the existence of the quark gluon
plasma (QGP). The detection of strangelets has long been a main subject in
the relativistic heavy ion collision (RHIC) experiments and the cosmic-ray
space experiments concerning with the search for anomalously heavy nuclei 
\cite{2,3,4}, and unfortunately, by now, it has not been found yet.

Recently, a very interesting anomalous event has been announced by the Alpha
Magnetic Spectrometer (AMS-01) Collaboration \cite{5}. Within more than four
million He events collected by AMS-01 detector, one that has $\frac ZA$ of $%
0.114\pm 0.01$, kinetic energy of $2.1$ $\mbox{GeV}$ and corresponds to the
flux of $5\times 10^{-5}(m^2\cdot sr\cdot \sec )^{-1}$ was found. This was a
doubly charged anomalous event and it strongly needed further explanation
from theoretical view of point whether the event denoted a small lump of
strange quark matter, i.e. strangelet, or only a nuclei \cite{5}.

This paper evolves from an attempt to give a possible explanation for this
doubly charged and heavy mass event. We will employ the quark mass
density-dependent (QMDD) model which was firstly suggested by Fowler, Raha
and Weiner \cite{6a} and then used by many authors to study the stability
and other properties of strange quark matter \cite{6b,6c,6d,6e}. In a series
of our previous papers \cite{7,8,9,10,11}, based on the Friedberg-Lee model 
\cite{12}, we extended the QMDD model to a quark mass density- and
temperature- dependent (QMDTD) model at finite temperature. The
thermodynamical properties and the stability of strangelets \cite{7,8,9},
the dibaryon system \cite{10} and the photo strange star \cite{13} have been
discussed by using QMDTD\ model. At zero temperature, the QMDTD model
reduces to the QMDD model. Since the AMS-01 experiment was flown on the
space shuttle Discovery during flight STS-91 (1998) in a 51.70 orbit at
altitudes between 320 and 390 Km, the temperature of cosmic space is very
low and we will neglect the temperature effect in our calculation.

According to QMDD model \cite{6a,6b,6c,6d,6e}, the masses of $u,d$ quarks
and strange quarks (and the corresponding anti-quarks) are given by

\begin{eqnarray}
m_q &=&{\frac B{3n_B}},\hspace{0.8cm}(q=u,d,\bar{u},\bar{d}),  \label{su1} \\
m_{s,\bar{s}} &=&m_{s0}+{\frac B{3n_B}},  \label{su2}
\end{eqnarray}
where $n_B$ is the baryon number density, $m_{s0}$ is the current mass of
the strange quark and $B$ is the vacuum energy density. One can proved that
the properties of strange matter in the QMDD\ model are nearly the same as
those obtained in the MIT bag model \cite{6c}.

At zero temperature, the particle number $N_i$ ($i=u,d,s$) reads

\begin{equation}
N_i=%
\displaystyle \int %
_0^{\sqrt{\mu _i^2-m_i^2}}\rho (k)dk,  \label{19zero}
\end{equation}
and the total energy of the strangelet is 
\begin{equation}
A=\sum_i%
\displaystyle \int %
_0^{\sqrt{\mu _i^2-m_i^2}}\sqrt{m_i^2+k^2}\rho (k)dk,  \label{20zero}
\end{equation}
where $\mu _i$ is the chemical potential of the quark and $\rho (k)$ is the
density of states. For a spherical cavity, $\rho (k)$ can be obtained by
multi-reflection theory \cite{14} or by numerical calculation \cite{15}. The
result is 
\begin{equation}
\rho (k)={\ 
{\displaystyle {d\left\{ \xi \cdot (kR)^3+\zeta \cdot (kR)^2+\eta \cdot (kR)\right\}  \over dk}}%
},  \label{a}
\end{equation}
where $R$ is the radius of the bag, 
\begin{equation}
\xi =\frac{2g}{9\pi },  \label{4}
\end{equation}
\begin{equation}
\zeta \left( \frac mk\right) =\frac g{2\pi }\left\{ \left[ 1+\left( \frac mk%
\right) ^2\right] \tan ^{-1}\left( \frac km\right) -\left( \frac mk\right) -%
\frac \pi 2\right\} ,  \label{5}
\end{equation}
and 
\begin{equation}
\eta \left( \frac mk\right) ={\frac g{2\pi }}\left\{ \frac 13+\left( {\frac k%
m}+{\frac mk}\right) \tan ^{-1}{\frac km}-{\frac{\pi k}{2m}}\right\} +\left( 
{\frac mk}\right) ^{1.45}{\frac g{3.42\left( {\frac{\displaystyle m}{%
\displaystyle k}}-6.5\right) ^2+100.}.}  \label{Our-C}
\end{equation}

The strangeness number $|S|$ of the strangelet reads 
\begin{equation}
|S|=N_s,  \label{21}
\end{equation}
the baryon number $N$ of the strangelet is 
\begin{equation}
N={\frac 13}(N_u+N_d+N_s)\hspace{0in},  \label{22}
\end{equation}
and the electric charge $Z$ of the strangelet is 
\begin{equation}
Z=\frac 23N_u-%
{\displaystyle {1 \over 3}}%
N_d\hspace{0in}-%
{\displaystyle {1 \over 3}}%
N_s.  \label{23}
\end{equation}
The stability condition of strangelets for the radius reads 
\begin{equation}
\frac{\delta A}{\delta R}=0.  \label{13}
\end{equation}

The chemical potentials of quarks $\mu _u,\mu _d,\mu _s$ of a strangelet
with fixed strangeness number, baryon number and electric charge could be
obtained by solving Eqs. (\ref{21}), (\ref{22}), (\ref{23}) and (\ref{13})
self-consistently. In the following, we limit our calculation in the area 
\begin{eqnarray}
Z &\geq &-N,  \label{25} \\
|S|+Z &\leq &2N  \label{26}
\end{eqnarray}
to keep the particle numbers of $u,$ $d,$ and $s$ quarks positive.

At zero temperature, the stability of a strangelet is determined by its
energy. Since the strangeness number $|S|$ and the electric charge $Z$ are
conserved in the strong process, a general expression of a two-body strong
decay for strangelet $Q(N,|S|,Z)$ can be written as

\begin{equation}
Q(N,|S|,Z)\rightarrow Q(N-1,|S|-|S_x|,Z-Z_x)+x(1,|S_x|,Z_x),  \label{28}
\end{equation}
with the energy balance of the corresponding reaction satisfies

\[
E(N,|S|,Z)>E(N-1,|S|-|S_x|,Z-Z_x)+m_{x,} 
\]
where $x$ stands for a baryon with strangeness number $S_x$ and electric
charge $Z_x$. For the weak process, the electric charge $Z$ is still
conserved but $\Delta S=\pm 1$, therefore, 
\begin{equation}
Q(N,|S|,Z)\rightarrow Q(N-1,|S|-|S_x|-1,Z-Z_x)+x(1,|S_x|,Z_x),  \label{30}
\end{equation}
if the energy balance of the corresponding reaction satisfies 
\begin{equation}
E(N,|S|,Z)>E(N-1,|S|-|S_x|-1,Z-Z_x)+m_{x.}  \label{31}
\end{equation}
A strangelet that decays via strong processes is called unstable strangelet
and the one could withstand strong decay and only decays via weak processes
is called metastable strangelet \cite{6}.

Our results are shown in Figure 1, 2 and Table 1. There are two parameters,
namely, bag constant $B$ and current mass of strange quark $m_{s0}$ in the
QMDD model. We fix $B_0=170$ $\mbox{MeV}$ $\mbox{fm}^{-3},$ $m_{s0}=150$ $%
\mbox{MeV}$ first. The stability of strangelets with fixed electric charge $%
Z=2$ and different baryon number $N$ and strangeness number $|S|$ is shown
in Figure 1, where open circles stand for the unstable strangelets and
filled circles stands for the metastable strangelets. The strangelet with
the energy most close to that of the anomalous event in AMS-01 detection ($%
\frac ZA$ $=0.114,$ $A=17.5$ $\mbox{u}$) is $Q(14,23,2)$, which is a
metastable strangelet and is represented by a filled square in Figure 1. The
strangelet $Q(14,23,2)$ is composed by $23$ strange quarks, $16$ up quarks
and $3$ down quarks. In Figure 2, we draw the curve of its energy as the
function of the radius. While the energy of the strangelet has its minimal
value, the corresponding radius is called the stable radius of the
strangelet, and the stable radius reads $1.992$ $\mbox{fm}$ for $Q(14,23,2)$.

One of the interested properties of the strangelet in the experimental
detection is its mean lifetime $\tau $. Employing the decay formula given by
Chin and Kerman\cite{17,10} 
\begin{equation}
1/\tau =[G^2\mu _s^5/192\pi ^3]\sin ^2\theta _cF(z),  \label{qing}
\end{equation}
with 
\begin{eqnarray}
F(z) &=&1-8z+8z^3-z^4-12z^2\ln z,  \label{fz} \\
z &=&\mu _u^2/\mu _s^2,  \label{z}
\end{eqnarray}
where the Cabibbo angle is given by $\sin \theta _c\simeq 0.22$, and
considering relativistic factor $\beta =0.462$ \cite{5}, we obtain that the
dynamical mean lifetime $\widetilde{\tau }$ for $Q(14,23,2)$ is $2.25\times
10^{-8}$ $\mbox{s}$, which is long enough for experimental detection.

Obviously, above result depends on the choice of parameters $B_{\text{ }}$%
and $m_{s0}$. In fact, the available choices of $B_{\text{ }}$and $m_{s0}$
are limited in an area called ''stability window'' which is shown in Ref. 
\cite{7} for QMDD\ model. To confirm our result, we choose 4 different
parameter pairs ($B,$ $m_{s0}$) in this window and other two pairs ($57.54$ $%
\mbox{MeV}$ $\mbox{fm}^{-3},$ $280$ $\mbox{MeV}$), ($396.93$ $\mbox{MeV}$ $%
\mbox{fm}^{-3},$ $150$ $\mbox{MeV}$) given by references \cite{6,18} to
calculate, and results are shown in Table 1

\[
\begin{tabular}{|c|c|c|c|c|c|c|}
\hline
$B$ $(\mbox{MeV}$ $\mbox{fm}^{-3})$ & $m_{s0}$ $(\mbox{MeV})$ & Metastable
Strangelet & Composition & $E$ $(\mbox{u})$ & $R$ $(\mbox{fm})$ & $%
\widetilde{\tau }$ $(\mbox{s})$ \\ \hline
170 & 150 & $Q(14,23,2)$ & $16u3d23s$ & 17.53 & 1.992 & $2.25\times 10^{-8}$
\\ \hline
180 & 150 & $Q(14,22,2)$ & $16u4d22s$ & 17.50 & 1.962 & $2.52\times 10^{-8}$
\\ \hline
190 & 150 & $Q(13,24,2)$ & $15u0d24s$ & 17.58 & 1.911 & $1.41\times 10^{-8}$
\\ \hline
220 & 75 & $Q(14,24,2)$ & $16u2d24s$ & 17.55 & 1.911 & $1.59\times 10^{-7}$
\\ \hline
57.54 & 280 & $Q(15,25,2)$ & $17u3d25s$ & 17.68 & 2.550 & $2.13\times
10^{-9} $ \\ \hline
396.93 & 150 & $Q(11,20,2)$ & $13u0d20s$ & 17.51 & 1.517 & $1.47\times
10^{-8}$ \\ \hline
\end{tabular}
\]

(Table 1).

We see from Table 1 that we can always find a metastable strangelet that
satisfies the condition $\frac ZA$ $=0.114$, although their compositions,
radiuses and dynamic mean lifetime are different. Therefore, we come to the
conclusion that the doubly charged anomalous heavy event detected by AMS-01
could attribute to a metastable strangelet, and this conclusion given by
QMDD\ model is not parameters dependent.

In summary, by means of the QMDD\ model, we give a possible explanation for
the doubly charged anomalous heavy event found by AMS-01 detector. It seems
a strangelet. This conclusion dose not depend on the choices of the model
parameters. Of course, we need more events to confirm the existence of the
strangelet and we hope AMS-02 could provide us more information of this
topic.

\section{Acknowledgment}

We thank professors Zuo-Xiu He and Cheng-Rue Qing for helpful discussions.
This work was supported in part by the NNSF of China under contract Nos.
10375013, 10247001 and 10235030, by the National Basic Research Program
2003CB716300 of China and the Foundation of Education Ministry of China
under contract 2003246005.

\section{Figure Captions}

Figure 1. The baryon number $N$ as a function of the strangeness number $S$
for unstable strangelets (open circles) and metastable strangelets (filled
circles) with electric charge number $Z=2$.

Figure 2. The energy of the strangelet $Q(14,23,2)$ as a function of radius $%
R$.

\section{Table Captions}

Table 1. The energy, stable radius, and dynamic mean lifetime of the
strangelets which has the energy around $17.5$ $\mbox{u}$ with different
parameters setting.

\begin{figure}[tbp]
\includegraphics[scale=0.5]{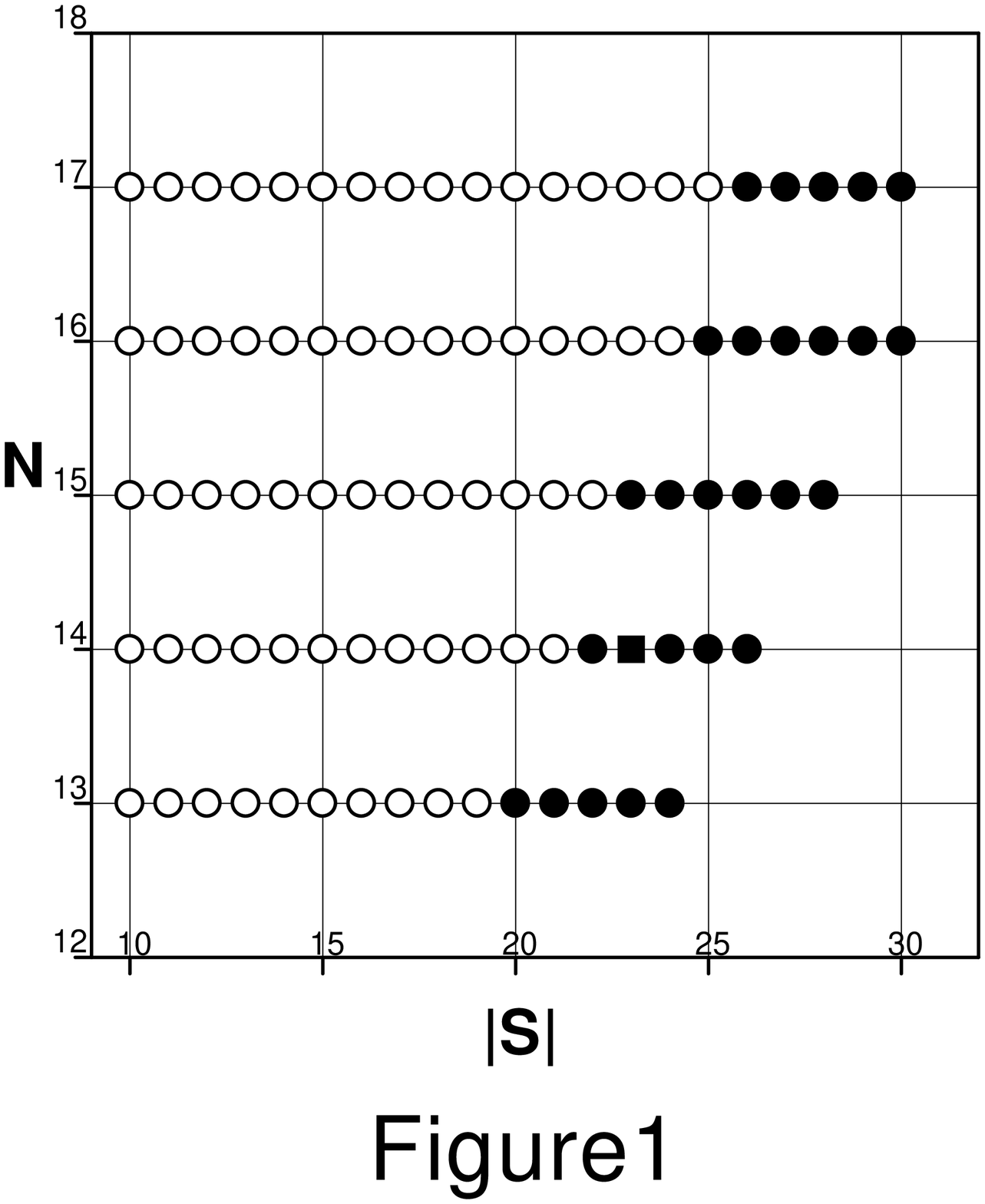}
\label{fig1}
\end{figure}

\begin{figure}[tbp]
\includegraphics[scale=0.5]{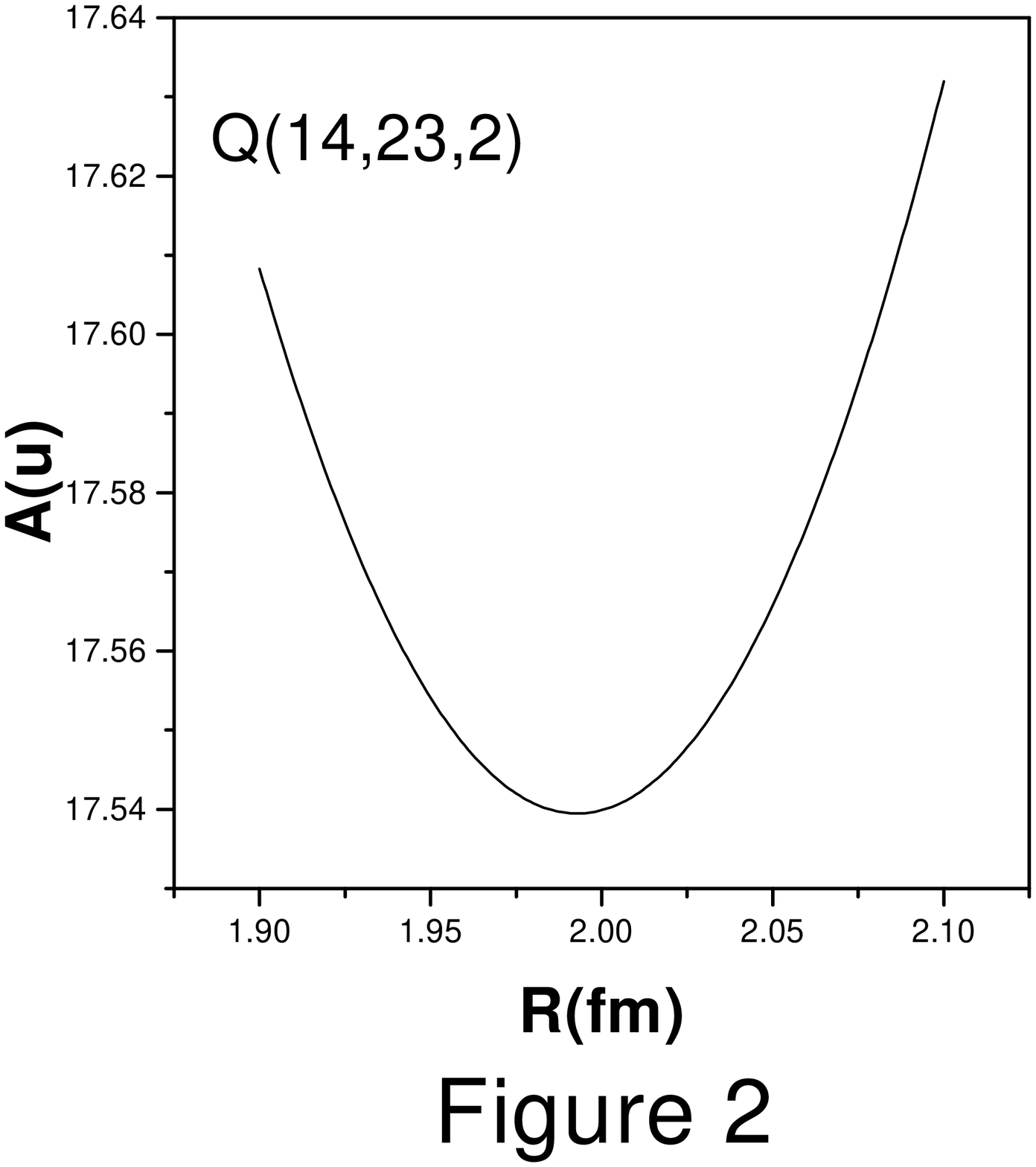}
\label{fig2}
\end{figure}

\end{document}